\documentclass{emulateapj}
\usepackage{apjfonts}

\journalinfo{The Astrophysical Journal, in press}
\slugcomment{Received 2007 May 26; accepted 2007 July 9}

\shortauthors{CAMILO ET AL.}
\shorttitle{VARIABLE SPECTRUM OF THE MAGNETAR XTE~J1810--197}

\begin{document}


\def\axp{XTE~J1810--197}

\title{The variable radio--to--X-ray spectrum of the magnetar XTE~J1810--197}

\author{F.~Camilo,\altaffilmark{1}
  S.~M.~Ransom,\altaffilmark{2}
  J.~Pe\~{n}alver,\altaffilmark{3}
  A.~Karastergiou,\altaffilmark{4}
  M.~H.~van~Kerkwijk,\altaffilmark{5}
  M.~Durant,\altaffilmark{6}
  J.~P.~Halpern,\altaffilmark{1}
  J.~Reynolds,\altaffilmark{7}
  C.~Thum,\altaffilmark{4}
  D.~J.~Helfand,\altaffilmark{1} 
  N.~Zimmerman,\altaffilmark{1}
  and I.~Cognard\altaffilmark{8} }

\altaffiltext{1}{Columbia Astrophysics Laboratory, Columbia University,
  New York, NY 10027.}
\altaffiltext{2}{National Radio Astronomy Observatory, Charlottesville,
  VA 22903.}
\altaffiltext{3}{Instituto de Radioastronom\'ia Millim\'etrica, E-18012
  Granada, Spain.}
\altaffiltext{4}{Institut de Radioastronomie Millimetrique, F-38406
  Saint Martin d'H\`eres, France.}
\altaffiltext{5}{Department of Astronomy and Astrophysics, University
  of Toronto, Toronto, ON M5S 3H4, Canada.}
\altaffiltext{6}{Instituto de Astrof\'isica de Canarias, E-38200 La
  Laguna, Tenerife, Spain.}
\altaffiltext{7}{Australia Telescope National Facility, CSIRO, Parkes
  Observatory, Parkes, NSW 2870, Australia.}
\altaffiltext{8}{Laboratoire de Physique et Chimie de l'Environnement,
  CNRS, F-45071 Orleans, France.}

\begin{abstract}
We have observed the 5.54\,s anomalous X-ray pulsar \axp\ at radio,
millimeter, and infrared (IR) wavelengths, with the aim of learning
about its broad-band spectrum.  At the IRAM 30\,m telescope, we
have detected the magnetar at $\nu = 88$ and 144\,GHz, the highest
radio-frequency emission ever seen from a pulsar.  At 88\,GHz we detected
numerous individual pulses, with typical widths $\sim 2$\,ms and peak
flux densities up to 45\,Jy.  Together with nearly contemporaneous
observations with the Parkes, Nan\c{c}ay, and Green Bank telescopes, we
find that in late 2006 July the spectral index of the pulsar was $-0.5
\la \alpha \la 0$ (with flux density $S_{\nu} \propto \nu^{\alpha}$)
over the range 1.4--144\,GHz.  Nine dual-frequency Very Large Array and
Australia Telescope Compact Array observations in 2006 May--September
are consistent with this finding, while showing variability of $\alpha$
with time.  We infer from the IRAM observations that \axp\ remains
highly linearly polarized at millimeter wavelengths.  Also, toward this
pulsar, the transition frequency between strong and weak scattering in
the interstellar medium may be near 50\,GHz.  At Gemini, we detected
the pulsar at $2.2\,\mu$m in 2006 September, at the faintest level
yet observed, $K_s=21.89\pm0.15$.  We have also analyzed four archival
IR Very Large Telescope observations (two unpublished), finding that
the brightness fluctuated within a factor of 2--3 over a span of 3
years, unlike the monotonic decay of the X-ray flux.  Thus, there is no
correlation between IR and X-ray flux, and it remains uncertain whether
there is any correlation between IR and radio flux.

\end{abstract}

\keywords{pulsars: individual (XTE~J1810--197) --- stars: neutron}

\section{Introduction}\label{sec:intro} 

Anomalous X-ray pulsars (AXPs) are neutron stars with long spin periods
and extremely strong surface magnetic fields, the decay of which is
largely responsible for their observed X-ray emission according to
the magnetar model \citep{dt92a}.  Of the dozen identified magnetars
(including soft-gamma repeaters in addition to AXPs), half have also
been detected at infrared (IR) wavelengths \citep[see][for a review of
magnetar properties]{wt06}.  While the IR fluxes of magnetars have been
observed to vary \citep[e.g.,][]{hvk04}, it is neither clear how these
variations relate to changes in observed X-ray flux, nor what is the
origin in detail of the IR radiation.

\axp\ is a transient AXP with period 5.54\,s, first detected
\citep{ims+04} when its X-ray flux increased $\sim 100$-fold compared
to the historical level maintained for at least 24 years \citep{hg05}.
Four years after the outburst, its X-ray flux has decreased nearly to
the quiescent level \citep{gh06}.  It was also detected in the IR in
2003 October \citep{irm+04}, and was observed 5 months later to be 60\%
fainter \citep{rti+04}.

\axp\ is the only magnetar known to emit radio waves \citep{hgb+05}.  This
emission is entirely pulsed \citep{crh+06,hcb+07}, is aligned in phase
with the X-ray pulsations \citep{ccr+07}, and is highly polarized, in some
respects being similar to that of ordinary young radio pulsars (Camilo et
al.\ 2007b; see also Kramer et al.\ 2007)\nocite{crj+07,ksj+07}.  However,
unlike in ordinary pulsars, the radio emission appears to be transient,
the average flux density varies intrinsically by factors of up to $\sim
3$ on approximately daily timescales, and the pulse profiles vary as
well \citep{crh+06}.  By early 2007, the average radio flux density was
a factor of $\sim 20$ lower than when pulsations were first observed 1
yr earlier \citep{ccr+07}.  Also, unlike the vast majority of ordinary
pulsars, \axp\ has a radio spectrum approximately consistent with being
flat ($\alpha \ga -0.5$; $S_\nu \propto \nu^{\alpha}$), over the large
frequency range $0.7 \le \nu \le 42$\,GHz \citep{crh+06}.

The radio window provides a new opportunity for learning about the
emission mechanism(s) of magnetars.  For instance, an extrapolation
of the radio spectrum of \axp\ exceeds its IR flux, so that the radio
and IR emission could have the same origin.  However, the shortest
radio wavelength at which a detection had been reported (7\,mm) is
3000 times larger than the IR wavelengths, so that coverage at shorter
radio wavelengths is desirable, also to provide further constraints on
the radio emission.  Variability in the IR and comparison with radio
variability would also be of great interest, but only two IR observations
have been reported, and they predate the discovery of radio pulsations.

With these objectives in mind, we report here on observations of \axp\
at the IRAM telescope, which has the capability to detect pulsars in the
1--3\,mm range, on simultaneous radio observations, and also on a new IR
observation at Gemini, along with analysis of archival IR observations
from the Very Large Telescope (VLT).

\section{Observations, Analysis, and Results}\label{sec:obs} 

\subsection{Millimeter wavelengths: IRAM}\label{sec:iram}

\begin{figure}
\begin{center}
\includegraphics[angle=0,scale=0.57]{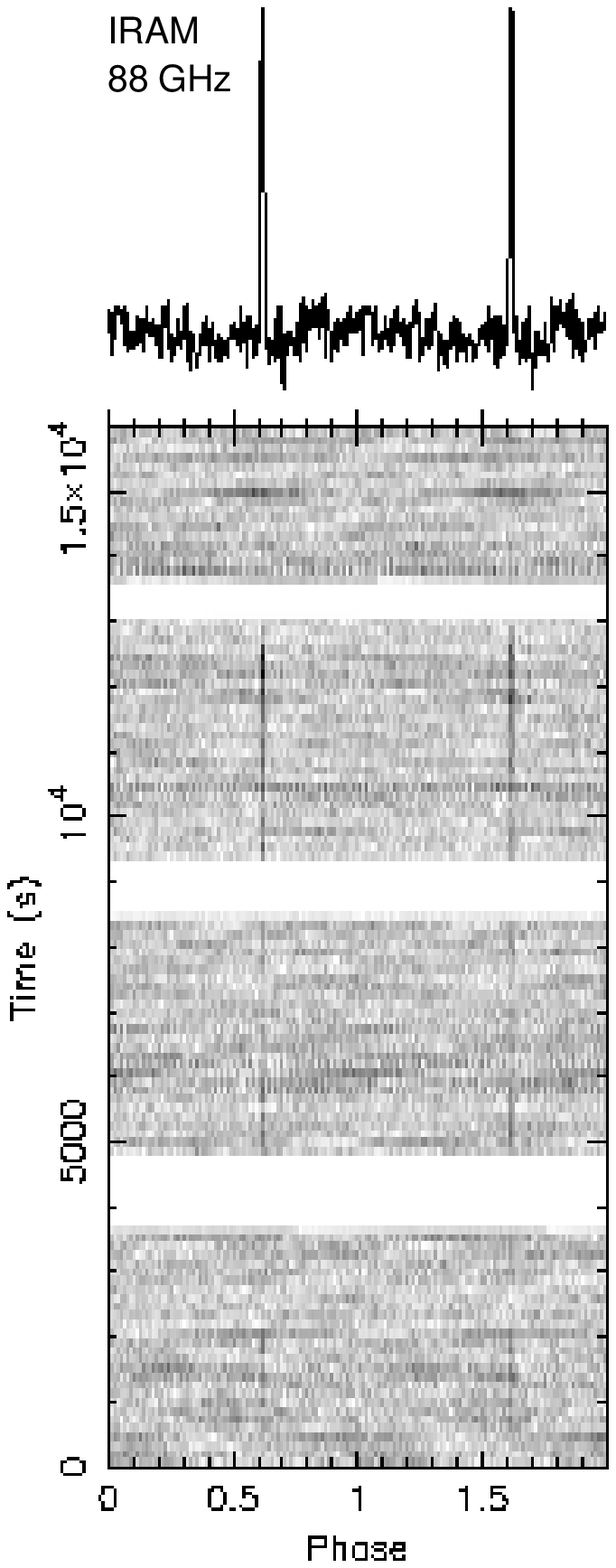}
\includegraphics[angle=0,scale=0.57]{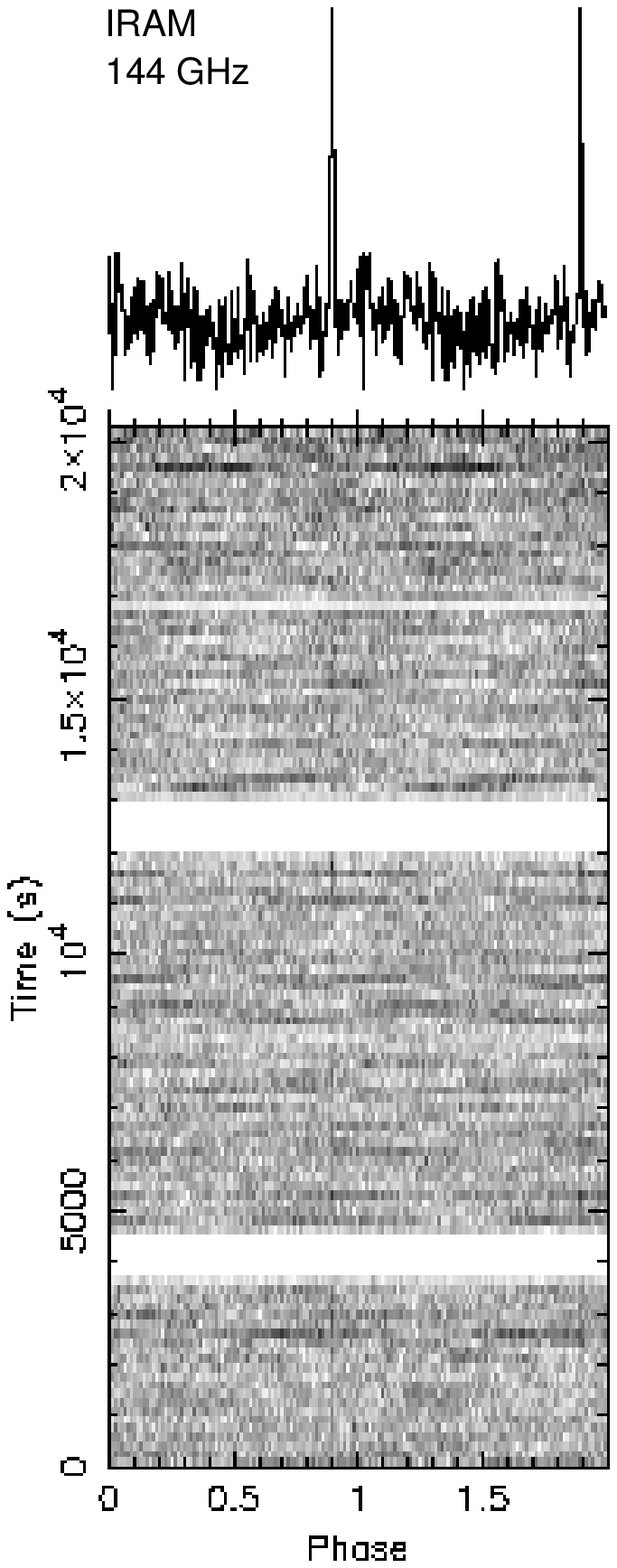}
\caption{\label{fig:iram}
\axp\ at IRAM at a central frequency of 88.5\,GHz on 2006 July 19
(\textit{left}) and 144\,GHz on July 20 (\textit{right}).  Only signal
from one of the polarizations (H) is folded in these plots.  The variation
with time of the signal strength (\textit{bottom plots}) is likely mainly
due to interstellar scintillation (see \S~\ref{sec:iram}).  The white
spaces correspond to times during which the telescope was not pointed
at \axp, and the pulse profiles displayed with 256 phase bins are shown
twice in each panel.
}
\end{center}
\end{figure}

We observed \axp\ at the IRAM 30\,m telescope in Pico Veleta, Spain,
during 2006 July 19--21\footnote{IRAM operates both a six-element
interferometer in France and the single dish in Spain; throughout
we refer to the IRAM telescope as a shorthand for the latter.}.
We recorded four intermediate frequencies (IFs) simultaneously.
These correspond to two linear polarizations (rotating with telescope
elevation, but parallel and perpendicular to the ground at the inputs
to the receivers, here designated H and V, respectively) at one of two
possible pairs of frequencies (cryostats AB, with receivers operating
at nominal frequencies of 100 and 230\,GHz, or CD, at 150 and 270\,GHz).
We used the AB receivers on July 19 and 21, and CD on July 20.  We tuned
all receivers in double side-band mode, with the IFs for the 100\,GHz
receivers centered at 1.5\,GHz with 1\,GHz bandwidth, and those for
all others centered at 4\,GHz with 2\,GHz bandwidth.  At the lowest
frequency used, we set the local oscillator to 88.5\,GHz, thereby
recording combined signals from $87\pm0.25$\,GHz and $90\pm0.25$\,GHz.
For the lowest frequency of the CD cryostat pair, signals were recorded
from $140\pm0.5$\,GHz and $148\pm0.5$\,GHz, for a central sky frequency
of 144\,GHz.  The other central frequencies used were 224 and 264\,GHz.
Dispersion is irrelevant at these frequencies, and we sampled the total
power IF signals every 0.5\,ms using a 16-bit analog-to-digital converter,
with the output recorded in VME modules using a custom configuration
built by J.P. and colleagues at IRAM.  The data stream containing
the sampled power from the 4 IFs and a time stamp, referenced to the
Observatory time standard and ultimately to UTC, was recorded to disk for
off-line analysis.  A typical observing session of 5\,hr consisted of
sets of pointing, focus, and flux-calibration scans (on Mars, Jupiter,
or Venus), interspersed with four scans on \axp, each about 1\,hr
in length, at elevations of $12\arcdeg \le \epsilon \le 39\arcdeg$.
Weather conditions were variable, and the sensitivity varied on each
night by up to 20\% at 88\,GHz and 50\% at 144\,GHz.

\begin{deluxetable}{cllll}
\tablewidth{0.94\linewidth}
\tablecaption{\label{tab:iram} Sensitivity of \axp\ observations at IRAM }
\tablecolumns{5}
\tablehead{
\colhead{Date}               &
\colhead{$S_{\rm sys}^{\nu_1\rm H}$} &
\colhead{$S_{\rm sys}^{\nu_1\rm V}$} &
\colhead{$S_{\rm sys}^{\nu_2\rm H}$} &
\colhead{$S_{\rm sys}^{\nu_2\rm V}$} \\
\colhead{(MJD)} &
\colhead{(Jy)}  &
\colhead{(Jy)}  &
\colhead{(Jy)}  &
\colhead{(Jy)}   
}
\startdata
53935 &  692--783  &  775--843  &  5640--7850  &  6430--8960  \\
53936 & 1780--2590 & 1640--2420 & 23000--53400 & 23400--55000 \\
53937 &  702--862  &  775--942  &  4290--10700 &  4940--12100 \\
54079 &  632--650  &  725--729  &  2130--2270  &  2400--2560  \\
54118 &  660--693  &  757--815  &  2650--2810  &  2800--3000  \\
54172 &  608--725  &  706--832  &  2250--2910  &  2440--2800  \\
54180 &  582--631  &  687--740  &  1900--2080  &  2080--2290  \\
54207 &  590--642  &  698--750  &  2700--3040  &  2810--3120
\enddata
\tablecomments{On MJD~53936, center frequencies were $\nu_1 = 144$ and
$\nu_2 = 264$\,GHz, respectively; on all other days, 88 and 224\,GHz.
For each frequency, we provide the range of measured system equivalent
flux density separately for the ``horizontal'' (H) and ``vertical''
(V) polarizations (see \S~\ref{sec:iram}).  These were all corrected for
air mass, measured hourly.  The bandwidths used were 1\,GHz at 88\,GHz
and 2\,GHz at other frequencies.  }
\end{deluxetable}

\begin{deluxetable}{ccclc}
\tablewidth{0.95\linewidth}
\tablecaption{\label{tab:radio} Radio and millimeter observations of \axp\ }
\tablecolumns{5}
\tablehead{
\colhead{Date}             &
\colhead{Time}             &
\colhead{Frequency $\nu$}        &
\colhead{$S_\nu$}          &
\colhead{Telescope }       \\
\colhead{(MJD/yymmdd)}     &
\colhead{(hr)}             &
\colhead{(GHz)}            &
\colhead{(mJy)}            &
\colhead{}                 
}
\startdata
53862.4/060507 & 0.2 &   1.4 &  $9.8\pm0.4$   & VLA        \\
53862.4/060507 & 0.2 &   8.5 &  $5.7\pm0.2$   & VLA        \\
53873.3/060518 & 0.6 &   1.4 &  $3.3\pm0.2$   & VLA        \\
53873.3/060518 & 0.6 &   4.9 &  $3.5\pm0.3$   & VLA        \\
53889.4/060603 & 0.7 &   1.4 &  $8.1\pm0.5$   & VLA        \\
53889.4/060603 & 0.7 &   4.9 &  $6.2\pm0.2$   & VLA        \\
53894.6/060608 & 7   &   1.4 &  $8.5\pm0.2$   & ATCA       \\
53894.6/060608 & 7   &   2.4 &  $8.5\pm0.2$   & ATCA       \\
53897.5/060611 & 0.7 &   1.4 & $10.4\pm0.5$   & VLA        \\
53897.5/060611 & 0.7 &   4.9 &  $7.5\pm0.2$   & VLA        \\
53905.5/060619 & 0.3 &   1.4 &  $8.0\pm0.9$   & VLA        \\
53905.5/060619 & 0.3 &   4.9 &  $8.0\pm0.2$   & VLA        \\
53906.1/060620 & 0.9 &  0.35 &   9.4          & GBT        \\
53935.5/060719 & 0.3 &   1.4 &   4.9          & Parkes     \\
53935.5/060719 & 0.6 &   1.4 &   2.6          & Parkes     \\
53935.6/060719 & 2.3 &   6.6 &   0.7          & Parkes     \\
53935.7/060719 & 0.4 &   1.4 &   1.9          & Parkes     \\
53935.9/060719 & 0.2 &   1.4 &   1.8          & Nan\c{c}ay \\
53935.9/060719 & 3.6 &  88.5 &   1.2          & IRAM       \\
53936.0/060720 & 1.5 &   9.0 &   0.7          & GBT        \\
53936.1/060720 & 0.9 &  19.0 &   1.0          & GBT        \\
53936.3/060720 & 0.7 &   1.4 &  $2.0\pm1.0$   & VLA        \\
53936.3/060720 & 0.7 &   4.9 &  $1.4\pm0.2$   & VLA        \\
53936.9/060720 & 4.7 & 144.0 &   1.2          & IRAM       \\
53937.9/060721 & 1.7 &  88.5 &   1.2          & IRAM       \\
53958.2/060811 & 0.7 &   1.4 &  $6.0\pm0.6$   & VLA        \\
53958.2/060811 & 0.7 &   4.9 &  $2.7\pm0.2$   & VLA        \\
53983.2/060905 & 0.7 &   1.4 &  $3.4\pm0.6$   & VLA        \\
53983.2/060905 & 0.7 &   4.9 &  $1.1\pm0.3$   & VLA        \\
53991.8/060913 & 0.3 &   1.4 &   1.6          & Nan\c{c}ay \\
53992.7/060914 & 0.2 &   1.4 &   1.6          & Nan\c{c}ay \\
54069.5/061130 & 1.1 &   1.4 &   0.4          & Nan\c{c}ay \\
54069.9/061130 & 0.6 &  0.35 &   1.6          & GBT        \\
54078.5/061209 & 0.7 &   1.4 &   0.7          & Nan\c{c}ay \\
54079.5/061210 & 3.6 &  88.5 &  $0.22\pm0.11$ & IRAM       \\
54118.4/070118 & 4.0 &  88.5 & $<0.1$         & IRAM       \\
54172.2/070313 & 4.5 &  88.5 & $<0.1$         & IRAM       \\
54180.2/070321 & 4.0 &  88.5 & $<0.1$         & IRAM       \\
54207.2/070417 & 3.2 &  88.5 &  $0.34\pm0.17$ & IRAM       \\
54208.2/070418 & 0.8 &   1.4 &   1.2          & Nan\c{c}ay
\enddata
\tablecomments{
We list here all our simultaneous dual-frequency observations, along with
selected others, ordered by date.  At IRAM, the frequency pairs were 88.5
and 224\,GHz, or 144 and 264\,GHz.  The pulsar was detected at only the
lower frequency of each pair (see \S~\ref{sec:iram}).  Unless specifically
given, the fractional flux density uncertainties are about 25\%. }
\end{deluxetable}

\begin{figure}
\begin{center}
\includegraphics[angle=0,scale=0.426]{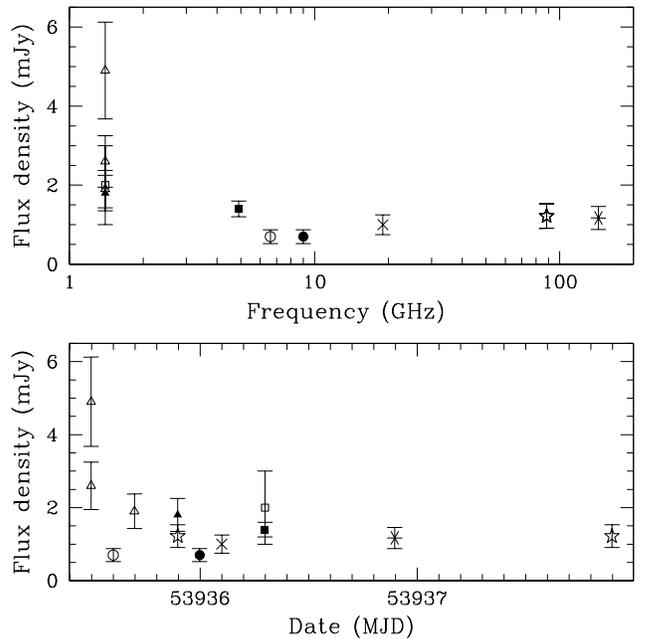}
\caption{\label{fig:fluxes}
Period-averaged flux densities of \axp\ across the frequency range
1.4--144\,GHz obtained over a period of 2.5 days (see \S\S~\ref{sec:iram}
and \ref{sec:radio}), plotted as a function of frequency (\textit{top})
and date (\textit{bottom}).  For each combination of frequency
and telescope we use a unique symbol that is common to both plots.
At 1.4\,GHz, multiple symbols represent the following data: Parkes
(\textit{open triangles}); Nan\c{c}ay (\textit{solid triangle}); VLA
(\textit{open square}).  With the exception of the VLA measurements
whose flux density uncertainties are recorded in Table~\ref{tab:radio},
the fractional uncertainties are $\sim 25\%$.  }
\end{center}
\end{figure}

Due to large changes in the atmospheric conditions at IRAM on timescales
of tens of seconds to several minutes, folding each \axp\ time-series
modulo the 5.54\,s pulse period (using the contemporary radio ephemeris;
\S~\ref{sec:radio}) resulted in a strongly varying off-pulse baseline
which greatly reduced the sensitivity to the detection of pulsations.
In addition, periodic interference contaminated the H polarization and
severely degraded the V polarization, greatly increasing the off-pulse
noise levels in the folded profiles.  We believe that these signals, at
50 and 100\,Hz, as well as many harmonics of 1\,Hz, are locally-generated
compressor-related interference.

In order to combat these issues, we filtered the data in two ways.  First,
to remove the low-frequency noise below $\sim 0.1$\,Hz (due primarily
to the changing atmosphere), we high-pass-filtered each time series
using a 3rd-order Bessel filter.  We then Fourier transformed the time
series and clipped the Fourier amplitudes of the strong 1\,Hz harmonics
as well as the prominent 50 and 100\,Hz signals.  Finally, we inverse
Fourier transformed the data to regenerate mostly interference-free
time series, which we then folded.  After filtering, the V polarization
sensitivity to pulsed signals remained significantly worse than that of
the H polarization.  The pulsar was not detected in the V polarization.
In the H polarization, the pulsar was clearly visible on both days at
88\,GHz (3.4\,mm) and at 144\,GHz (2.1\,mm; Fig.~\ref{fig:iram}), and
was not detected at higher frequencies.

We determined the period-averaged flux density by measuring the area under
the profiles and dividing by the pulsar period.  This was converted to an
absolute Jansky scale using the hourly flux-calibration scans, from which
the system equivalent flux density $S_{\rm sys}$ was calculated for each
frequency and polarization.  The off-pulse profile rms then corresponds
to $S_{\rm sys} (B T)^{-1/2}$ for a bandwidth $B$ and an integration
time $T$, where $S_{\rm sys}$ is corrected for air-mass attenuation
$\exp(-\tau/\sin\,\epsilon)$, and $\tau$ is the measured zenith opacity.
The ranges of $S_{\rm sys}$ obtained for each observation are listed
in Table~\ref{tab:iram}.

\begin{figure}
\begin{center}
\includegraphics[angle=0,scale=0.90]{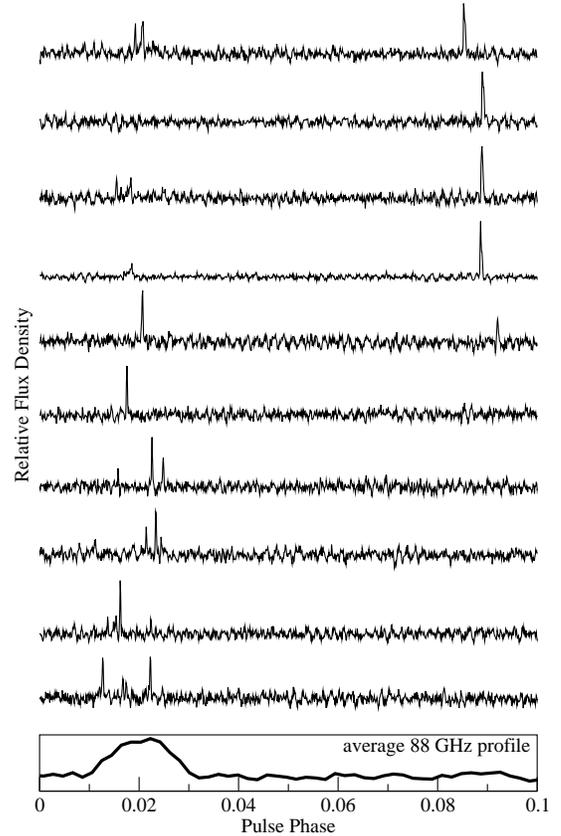}
\caption{\label{fig:sp}
Single pulses of \axp\ at 88\,GHz detected at IRAM on 2006 July 19.
The 10 single pulses, displayed with 0.5\,ms resolution, were chosen
from among the brightest in our data and are time-ordered (though
not consecutive) from top to bottom.  At bottom is the average pulse
profile on this day, displayed with 11\,ms resolution (see also
Fig.~\ref{fig:iram}).  Only 10\% of pulse phase is shown in order to
emphasize the sharpness of the sub-pulse structure (with typical width
$\sim 2$\,ms); no emission was detected from other phases.  Emission
such as seen in the top five pulses trailing the main component(s)
by $\sim 0.07$ in phase was among the brightest in the dataset, but
was infrequent enough that it did not contribute significantly to the
average profile.  The brightest pulse detected is shown fourth from
the top, with a signal-to-noise ratio of 45 corresponding to a peak
flux density of approximately 45\,Jy.  As in Figs.~\ref{fig:iram} and
\ref{fig:profs}, only signal from one polarization is displayed here.
The high-frequency noise readily apparent in the baseline of the profiles
is due to imperfectly excised interference.  }
\end{center}
\end{figure}

It is clear from Figure~\ref{fig:iram} that the observed flux density
can vary greatly with time.  Changing weather conditions and elevation
account for only up to 10\% of this variation in the July 19 data,
displayed in the left panel of the figure (see Table~\ref{tab:iram}).
A great portion of the variation is most likely caused by interstellar
scintillation \citep[see][]{crh+06}.  As an average, we estimate
the flux density to have been 1.2\,mJy at both 88 and 144\,GHz,
with a fractional uncertainty on the absolute values of about 25\%
(see Fig.~\ref{fig:fluxes} and Table~\ref{tab:radio}).  In order to
calculate approximate flux density limits for non-detections, we assume
a threshold signal-to-noise ratio of 5 and a pulse duty cycle of 2\%.
For example, on July 19 we obtain $S_{224} \la 0.9$\,mJy.

\begin{figure}
\begin{center}
\includegraphics[angle=0,scale=0.45]{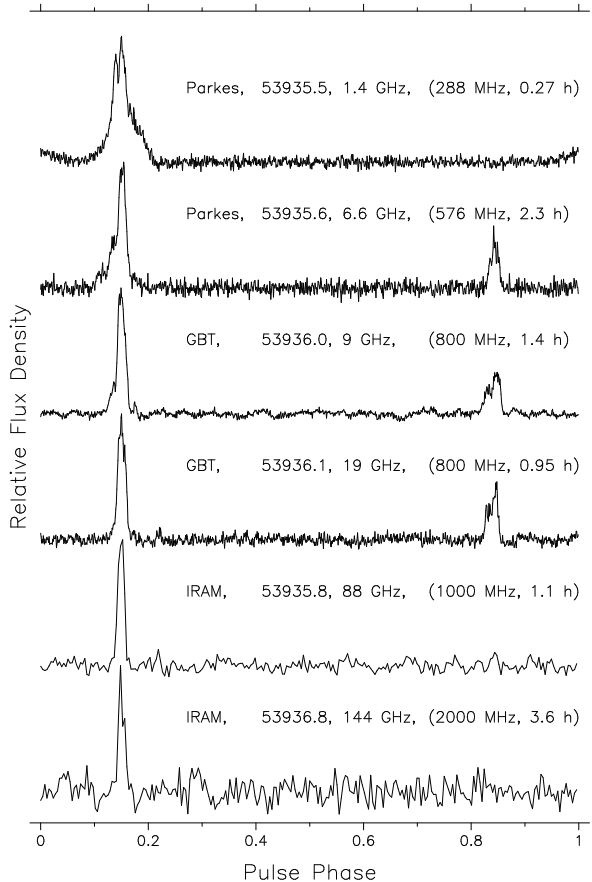}
\caption{\label{fig:profs}
Average pulse profiles of \axp\ obtained at frequencies spanning
1.4--144\,GHz over a period of 1.5 days, at Parkes, GBT, and IRAM.
Each profile is labeled by the telescope used, the date (MJD) of the
observation, the central frequency and, in parentheses, the bandwidth
and integration time used.  IRAM profiles include data from only one
polarization (see \S~\ref{sec:iram}), and are displayed with 256 phase
bins.  Parkes and GBT profiles have 1024 bins.  Profiles were aligned
by eye such that the peak pulse component arrives near phase 0.15.
The FWHM of these profiles are, in order of increasing frequency, 4.2,
2.2, 1.7, 1.6, 1.3, and 1.3\% of the pulse period.
}
\end{center}
\end{figure}

The observed flux density was so great on July 19 toward the end of the
third scan (Fig.~\ref{fig:iram}) that we were able to detect numerous
individual pulses from \axp\ at 88\,GHz.  On this day we detected single
pulses with signal-to-noise ratio $>4$ from about 15\% of all rotations
of the neutron star (Fig.~\ref{fig:sp}).  The largest pulses had a peak
flux density of $\sim 45$\,Jy, comparable to the strongest celestial
sources known at 3\,mm with the exception of the Sun, Jupiter, and Venus
(although only for $\sim 1$\,ms out of every $\sim 5$\,s).  On July 21
we also detected single pulses at 88\,GHz.

Following these detections, we monitored the pulsar at 88/224\,GHz on
five occasions between 2006 December and 2007 April, using identical
observing parameters.  These latter observations were done in winter and
early spring, under better weather conditions (Table~\ref{tab:iram}).
The pulsar was detected on two of these occasions, at 88\,GHz in the
H polarization, in December and April, with $S_{88} = 0.2$--0.3\,mJy
(see Table~\ref{tab:radio}).  In 2007 April, $S_{224} \la 0.4$\,mJy.

\subsection{Radio: Parkes, Nan\c{c}ay, GBT, VLA, and ATCA}\label{sec:radio}

Because the intrinsic flux density and pulse profile of \axp\ vary
on $\sim 1$ day timescales, determination of a spectral index ideally
requires simultaneous multi-frequency observations.  We therefore observed
at Parkes, Nan\c{c}ay, the Green Bank Telescope (GBT), and the Very
Large Array (VLA) on 2006 July 19--20, nearly simultaneously with IRAM.
Table~\ref{tab:radio} lists these and all dual-frequency observations at
the VLA and the Australia Telescope Compact Array (ATCA).  The methods
used have been described elsewhere (see Camilo et al.\ 2006 for Parkes
and GBT; Camilo et al.\ 2007a for Nan\c{c}ay and VLA; Helfand et al.\
2007 for ATCA). \nocite{crh+06,ccr+07,hcb+07}

The 1.4\,GHz flux density decreased over a period of a few hours
shortly before the first IRAM observation (Fig.~\ref{fig:fluxes}
and Table~\ref{tab:radio}).  The average pulse profiles during
these different flux states were identical (we show the first one in
Fig.~\ref{fig:profs}), unlike an observation in 2006 September with
sudden simultaneous flux and profile changes \citep{ccr+07}.

The nominal flux density of the pulsar at 19\,GHz was slightly greater
than at 9\,GHz (see Fig.~\ref{fig:fluxes}).  However, the observed
flux at $\nu \ga 9$\,GHz varies on short timescales due to interstellar
scintillation (Camilo et al.\ 2006\nocite{crh+06}; Fig.~\ref{fig:iram}),
and this may bias somewhat those flux densities obtained here (i.e.,
1\,hr does not correspond to many scintles).

\begin{figure}
\begin{center}
\includegraphics[angle=0,scale=0.426]{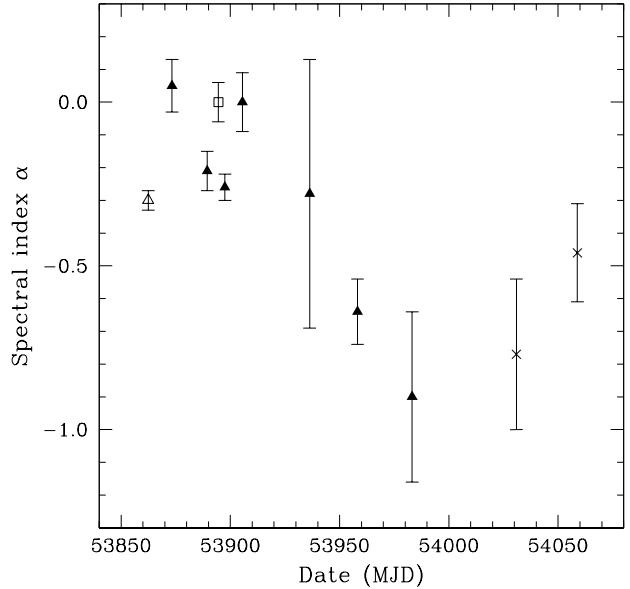}
\caption{\label{fig:alpha}
Spectral indices for \axp\ determined from simultaneous dual-frequency
VLA (\textit{triangles}) or ATCA (\textit{square}) observations, or
near-simultaneous GBT data (\textit{crosses}).  For each simultaneous
observation the lowest frequency used was 1.4\,GHz.  For the VLA the
second frequency was 8.5\,GHz for the first point plotted, and 4.9\,GHz
for all others (see Table~\ref{tab:radio}).  The point with the largest
error bar was obtained on 2006 July 20.  The first GBT point is based
on flux density data obtained at 1.9 and 9\,GHz over a 4\,hr interval.
The last point is based on observations at 0.8, 1.9, 5, and 9\,GHz,
obtained over 7\,hr.  Error bars are $1\,\sigma$.
}
\end{center}
\end{figure}

Using simultaneous 1.4 and 4.9\,GHz observations on 2006 July 20, we
measured a spectral index $\alpha=-0.3\pm0.4$.  The large uncertainty is
due to the uncertain 1.4\,GHz flux density, the smallest ever measured
at the VLA for the magnetar.  On a total of nine occasions we have
used accurately flux-calibrated simultaneous dual-frequency data to
determine the spectral index of \axp, shown in Figure~\ref{fig:alpha}.
From this, it is apparent that the radio spectrum, averaged over all its
pulse profile components, varies, and there is a hint that since late
2006 July it might have become steeper (at least in the 1.4--4.9\,GHz
range).  This appears to be supported by estimates of $\alpha$ obtained
from near-simultaneous multi-frequency pulsed flux measurements at GBT:
those shown in Figure~\ref{fig:alpha} for two epochs in 2006 October and
November have $\alpha \la -0.5$, while before mid-2006, $\alpha >-0.5$
\citep{crh+06}.  We have also extended the range of frequencies over which
the magnetar has been detected, with two GBT observations at 0.35\,GHz.
Together with observations at other frequencies within half a day of
these (Table~\ref{tab:radio}), we obtain $\alpha = 0$ in 2006 June,
but $\alpha = -1$ in 2006 November.  This further supports the notion
that the spectrum of \axp, while variable with $\alpha$ ranging between
approximately 0 and --1, may have become generally steeper after mid-2006.

\subsection{Infrared: Gemini and VLT}\label{sec:ir}

We obtained a near-IR $K_s$-band ($2.15\,\mu$m) observation of \axp\
at Gemini-North on 2006 September 14.  We used the adaptive optics
(AO) system Altair with the near-IR imager NIRI \citep{hji+03}.
With this configuration, the $1024\times1024$ pixel Aladdin InSb array
covers $22\times22{\rm\,arcsec^2}$ at $21.9{\rm\,mas\,pixel^{-1}}$.
For photometric calibration, as well as the astrometric analysis described
in \cite{hcb+07}, we also analyzed $K_s$-band observations taken on 2003
September 18 with NIRI on Gemini without the AO system (for which the
detector covers $2\times2{\rm\,arcmin^2}$ at $117{\rm\,mas\,pixel^{-1}}$;
19\,minute exposure time).

For comparison, we analyzed observations taken at the VLT using
the AO system NAOS with the CONICA camera \citep{lhb03,rlp+03}.
NAOS-CONICA also uses a $1024\times1024$ Aladdin detector, covering
$27\times27{\rm\,arcsec^2}$ at $27.0{\rm\,mas\,pixel^{-1}}$.
These observations were obtained in 2003 October \citep{irm+04}, 2004
March \citep{rti+04}, and 2004 September (on two nights, previously
unpublished).  We summarize in Table~\ref{tab:ir} the IR observations
of \axp\ analyzed here in detail.

\begin{deluxetable}{lcll}
\tablewidth{0.86\linewidth}
\tablecaption{\label{tab:ir} Infrared observations of \axp\ }
\tablecolumns{4}
\tablehead{
\colhead{Date}             &
\colhead{Exposure}         &
\colhead{$K_s$ magnitude}  &
\colhead{Refs. }           \\
\colhead{(MJD/yymmdd)}     &
\colhead{(minutes)}            &
\colhead{}                 
}
\startdata
52920.05/031008 & 32  & $20.9  \pm 0.2 $ &  (1) \\
---             & --- & $20.8  \pm 0.1 $ &  (2) \\
53078.37/040314 & 36  & $21.21 \pm 0.14$ &  (1) \\
---             & --- & $21.36 \pm 0.07$ &  (3) \\
53258.99/040910 & 36  & $20.9  \pm 0.2 $ &  (1) \\
53259.99/040911 & 60  & $20.82 \pm 0.16$ &  (1) \\
53992.21/060914 & 38  & $21.89 \pm 0.15$ &  (1)
\enddata
\tablecomments{Two entries are provided for each of the first two
observations, corresponding respectively to our analysis, and to the
original published analysis.  The first four observations listed
were obtained at the VLT, while Gemini was used for the last one.
Dates correspond to observation start times.  All observations used
adaptive optics, and the magnitudes listed are on the 2MASS system (see
\S~\ref{sec:ir}).  Uncertainties are all given at the $1\,\sigma$ level. }
\tablerefs{(1) This work; (2) \cite{irm+04}; (3) \cite{rti+04}. }
\end{deluxetable}

All observations were taken in a similar way, with images taken at
dithered positions, and each image consisting of one or more co-added
exposures, with the counts in each exposure determined from the difference
between series of read-outs before and after the actual integrations.
We reduced all observations in an identical fashion, using the Munich
Image Data Analysis System (MIDAS) or the Image Reduction and Analysis
Facility (IRAF).  We corrected for pixel-to-pixel sensitivity variations
using sky flats constructed from the images themselves, aligned the
images to integer pixel boundaries, and took averages.  The 2003 October
data had unrepeated horizontal ``banding'' that was not removed by the
flat-fielding process.  We removed this feature using the median value
for each data row before stacking images. In \cite{hcb+07}, we used the
2004 and 2006 observations in Table~\ref{tab:ir} to confirm that the
IR counterpart has the same proper motion as that measured from radio
VLBA observations.

\begin{figure}
\begin{center}
\includegraphics[angle=0,scale=0.56]{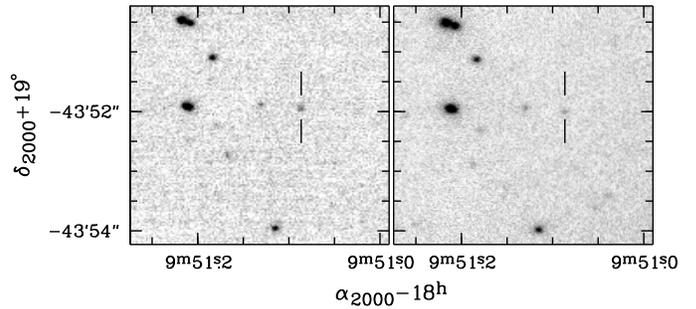}
\caption{\label{fig:ir}
Adaptive optics images of \axp\ at $2.2\,\mu$m taken with the VLT in
2004 March ({\em left}) and with Gemini in 2006 September ({\em right}).
The tickmarks are centered on the accurate VLBA position.  Between the
two epochs, the infrared counterpart has faded (see \S~\ref{sec:ir}
for details).  }
\end{center}
\end{figure}

In Figure~\ref{fig:ir} we compare the final VLT image from 2004 March with
our new Gemini image from 2006 September.  Over this 2.5\,yr period,
the IR counterpart has moved about one pixel to the south relative
to local stars due to its proper motion, and it has faded.  For a
quantitative analysis of all the images, we used the DAOPHOT~II package
\citep{ste87}, running inside MIDAS and IRAF, to derive instrumental
magnitudes by fitting a model point-spread function (PSF) for stars on
the average images.  Following the recommendations of \citet{ste87},
the PSF was derived from brighter stars in an iterative fashion, where
in each iteration nearby fainter stars are removed with the current
best model. For a well-matched model PSF, with minimal residuals, we
chose a Lorentzian analytical base, and linear dependence of the shape
on position.  We determined uncertainties on the instrumental magnitudes
by adding fake stars of a range of brightness to the images, re-measuring
those, and determining the rms deviations.  We found that these numbers
were very similar to the measurement uncertainties returned by DAOPHOT.

We calibrated our instrumental magnitudes relative to the 2MASS (Two
Micron All Sky Survey) catalog \citep{scs+06} in two steps.  First,
we derived offsets for 2MASS stars measured in the wider 2003 September
NIRI image.  We found that a constant offset gave a good fit, with the
rms residual of 0.06\,mag being only slightly larger than the typical
2MASS measurement error of typically $\sim 0.05\,$mag for the 21 stars
we used.  Thus, the photometry of this image is tied to the 2MASS system
to about 0.01\,mag.

We then transferred the photometry to the AO images using fainter stars.
For this step, we found that for most images the residuals around the mean
of instrumental minus calibrated magnitudes had rather large rms, of 0.02,
0.08, 0.06, 0.06, and 0.10\,mag respectively for each the five data sets
(see Table~\ref{tab:ir}), with no obvious dependencies on brightness,
position, or proximity to the guide star used.  Since most of these rms
residuals are much larger than the measurement uncertainties, we use
them as a measure of the calibration uncertainty for any given star.
Likely, the large scatter reflects the general difficulty of doing
reliable photometry on AO-corrected images.  Indeed, the largest scatter
is found for the 2004 March VLT and the 2006 Gemini data, which had the
best AO correction, while the smallest is for the 2003 October VLT data,
which had the worst correction.

The $K_s$ magnitudes thus obtained for \axp\ are listed in
Table~\ref{tab:ir}, with measurement and calibration uncertainties added
in quadrature (all uncertainties here are given at the $1\,\sigma$
confidence level).  The values for 2003 October and 2004 March are
consistent with those reported by \citet{irm+04} and \citet{rti+04},
respectively, but our uncertainties are larger by a factor of 2.
This cannot be due solely to the fact that we included the calibration
uncertainty, since for the two observations concerned this is smaller
than the measurement uncertainty for the faint target.  In order to
verify our estimates of the uncertainties, we checked the measurements
of three other, relatively faint stars (at $\approx 2.0''$ SSE,
$1.7''$ ENE, and $2.8''$ N of the target; the former two are visible
in Fig.~\ref{fig:ir}).  For the five observations, their magnitudes are
[20.35(12), 20.39(11), 20.30(12), 20.33(11), 20.39(10)], [20.30(11),
20.33(11), 20.23(11), 20.27(10), 20.40(10)], and [21.5(4), 20.94(14),
21.0(2), 21.0(2), 21.01(11)], where the uncertainties again include the
calibration uncertainties.  The uncertainty estimates are comparable to
the scatter between the magnitudes, suggesting that they are reasonable
(if perhaps slightly high).

Independent of the details of the uncertainties, our measurements show
that the IR brightness of \axp\ did not vary greatly in 2003 and 2004,
by $\la 50$\%, declining more sharply only by 2006.  Overall, these
(sparsely sampled) observations do not show a clear trend.  We also
searched for intra-night variations in the 2004 March data, by considering
the observation in two separate 18\,minute segments, finding a magnitude
difference of $0.11 \pm 0.23$.  At the $\sim 1$\,hr or $\sim 1$\,day
timescales, we thus have found no evidence for variations at the $\ga
20\%$ level.

Finally, using the $K_s$-band zero point of \cite{cwm03}, the Gemini
measurement in 2006 September corresponds to a flux density of
$1.17\pm0.17\,\mu$Jy.  If $A_V=3.6$ \citep{hg05}, then $A_K = 0.40$
\citep{sfd98}, so that the de-reddened flux is 1.45 times larger,
$1.69\pm0.25\,\mu$Jy.


\section{Discussion}\label{sec:disc} 

The remarkable detection of \axp\ at frequencies of 88 and 144\,GHz
with IRAM is a record among pulsars \citep[see][for previous
highest-frequency detections]{mkt+97}.  This magnetar has now been
detected at radio--millimeter wavelengths spanning a factor of 400.
The spectrum over this expanse is flat compared to that of most pulsars
\citep[$\alpha \approx -1.6$, measured over a narrower frequency range;
see][]{lylg95}, but it is difficult to make a definitive quantitative
statement: (1) as for ordinary pulsars \citep[e.g.,][]{kkg+03}, individual
phase components of the pulse profile may have differing spectra; (2)
at high frequencies ($\ga 9$\,GHz) there is considerable variability in
measured flux density due to interstellar scintillation; (3) the intrinsic
flux varies with time; (4) the spectrum also apparently varies with time
(see Fig.~\ref{fig:alpha}); and (5) the intrinsic spectrum need not be
represented by a single power law.

The only simultaneous multi-frequency detections among those summarized
in Figure~\ref{fig:fluxes} were the ones at the VLA (1.4 and 4.9\,GHz),
and 1.4\,GHz at Nan\c{c}ay along with 88\,GHz at IRAM.  The spectral
index obtained from the first pair is $\alpha=-0.3$, while from the
second, $\alpha=-0.1$, in both cases with substantial uncertainties
and also suffering from some of the problems noted above.  Attempts to
infer $\alpha$ from other measurements collected within a period of
2.5 days (see Fig.~\ref{fig:fluxes}), also suffered from such problems.
(See also Fig.~\ref{fig:alpha}.)  With the foregoing caveats in mind,
we summarize the situation thus: before 2006 August, the pulse-averaged
spectrum of \axp\ over the range 1.4--144\,GHz could be usefully described
by a single spectral index in the range $-0.5 \la \alpha \la 0$, with
some time variability observed.  Since then, the spectrum has apparently
steepened, with $-1.0 \la \alpha \la -0.5$.

In retrospect, the flux of the pulsar started changing dramatically
around mid 2006 July, near the time of the first IRAM observations.
This was accompanied by a change in the character of pulse profile
variations, as well as large torque variations \citep[see][]{ccr+07}.
The flux density at 1.4\,GHz by early 2007 was generally quite low
($\la 0.5$\,mJy --- although fluctuations continue: see the last entry
in Table~\ref{tab:radio}).  Together with a moderate spectral index
of $\alpha \la -0.3$, this can account for the non-detections at $\ge
88$\,GHz in 2007 (Table~\ref{tab:radio}).

The average pulse profiles of \axp\ observed during 2006 July 19--20
and spanning 1.4--144\,GHz are shown in Figure~\ref{fig:profs}.
The profiles at 6.6--19\,GHz are very similar, while in all 1.4\,GHz
observations on this date the ``precursor'' component is absent (phase
0.85 in the figure).  This is unexpected, because the Parkes 1.4\,GHz
observations bracketed in time that at 6.6\,GHz, ensuring that the
differences observed between the pulse profiles were not caused by a
global change in the profile.  Rather, the absence of the precursor at
1.4\,GHz may point on this day either to an extremely positive spectral
index for that component ($\alpha > 1$), or to absorption of the 1.4\,GHz
radiation along the line of sight at the location in the magnetosphere
corresponding to this pulse phase.

The absence of the precursor at 88\,GHz may be explained by a spectrum
over 19--88\,GHz steeper than that of the main component by $\Delta \alpha
\la -0.7$.  In making this estimate we assumed that roughly half of the
precursor power would be present in each of the H and V polarizations.
This follows from a computation of the polarimetric Stokes parameters
as they would appear for the IRAM observations, under the assumption
that the precursor was polarized at millimeter wavelengths as it was
earlier at 1.4\,GHz \citep[see][]{crj+07}.  A similar computation for
the main profile component using the (different) absolute position angle
of linear polarization observed at 1.4--8.4\,GHz \citep{crj+07}, shows
that the vast majority of its power in the IRAM observations should be
present in the H polarization, as is indeed the case (\S~\ref{sec:iram}).
This provides evidence that \axp\ remains highly linearly polarized at
wavelengths as short as 2\,mm.  For comparison, some ordinary pulsars
appear to depolarize significantly with decreasing wavelength, down to
the limit of the observations at 1\,cm \citep{xkj+96}.  However, separate
profile components can evolve differently with wavelength.  \citet{kjm05}
consider a model where orthogonal polarized modes have different spectra,
leading naturally to some highly polarized components with flat spectra,
down to 10\,cm wavelength.  Also, several young pulsars remain highly
polarized down to the limit of the observations at 3.5\,cm \citep{jkw06}.
It is therefore possible that in this respect \axp\ differs from ordinary
young pulsars more in having a flat spectrum all the way down to 2\,mm,
which enables us to make the measurements, than in remaining highly
polarized at those wavelengths \citep[see also][]{crj+07}.  In any case,
these two features may not be independent.

As seen clearly in Figure~\ref{fig:iram}, the flux density of \axp\
observed at millimeter wavelengths through a bandwidth of 1--2\,GHz varies
by factors of a few on timescales of $\sim 1$\,hr.  Most of this variation
is likely caused by interstellar scintillation (\S~\ref{sec:iram}).
The surprise is not that the received flux varies, but that it varies
so relatively little.  Qualitatively, the flux modulation for these
IRAM observations is much smaller than that observed at 42\,GHz and more
comparable to that observed at 14--19\,GHz \citep[cf.][]{crh+06}.  Perhaps
for this object, the transition between strong and weak interstellar
scattering \citep[at which the flux modulation peaks; see, e.g.,][]{lk05}
is about 50\,GHz.  Detailed analysis of scintillating behavior for \axp\
will be presented by Ransom et al. (in preparation).

The single pulses seen from \axp\ at 88\,GHz on 2006 July 19
(Fig.~\ref{fig:sp}) appear qualitatively similar to those observed at 2
and 42\,GHz about 2 months earlier \citep{crh+06}.  In particular, they
are narrow (a few ms), wander about with no obvious phase coherence
from rotation to rotation, and gradually build up the much wider
average profile.  Also, some particularly strong pulses are emitted from
rotational phases at which the average pulse is weak or not detectable,
implying significantly different pulse energy distributions at different
phases.  A detailed treatment of single-pulse behavior from \axp\ will
be presented elsewhere.

\begin{figure}
\begin{center}
\includegraphics[angle=0,scale=0.426]{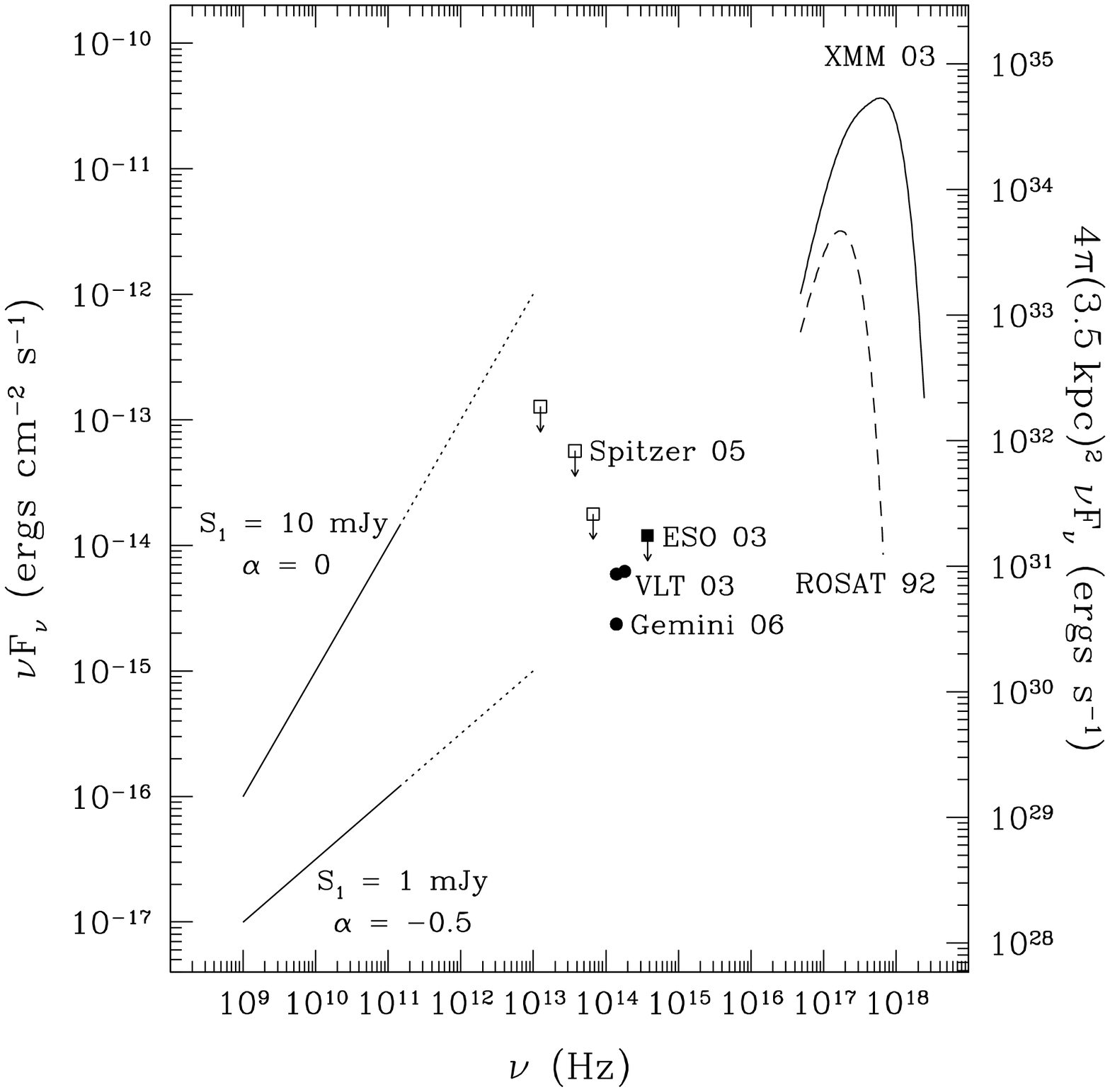}
\caption{\label{fig:sed}
Energy flux (\textit{left axis}) versus frequency for \axp.  The {\em
ROSAT\/} and {\em XMM\/} curves are fits to X-ray measurements obtained
in 1992 and 2003 September, respectively, and approximately bracket
the observed flux range of the source \citep[see][]{hg05}.  In the
optical/IR range, squares with arrows denote upper limits, obtained
at $I$ band \citep[][``ESO~03'']{irm+04} and mid-IR wavelengths
\citep[][``Spitzer~05'']{wkh07}.  The IR detections \citep[][and
\S~\ref{sec:ir}; \textit{circles}]{irm+04} bracket the range observed
(see Table~\ref{tab:ir}).  All IR--optical--X-ray measurements have been
corrected for interstellar absorption.  At radio frequencies, we have
drawn two solid lines between 1 and 144\,GHz, corresponding to a flat
spectrum with a flux density of 10\,mJy (\textit{upper}), and a spectrum
with $\alpha = -0.5$ and 1\,GHz flux density of 1\,mJy (\textit{lower}).
These spectra crudely bracket the range of variability displayed by
the pulsar in 2006 (see text).  In each case we also extrapolated the
spectra to $10^{13}$\,Hz (\textit{dotted lines}).  Very importantly,
the only data point in the figure that is more or less contemporaneous
with any of the radio detections is the ``Gemini~06'' IR observation (see
\S~\ref{sec:ir}), so that interpretation of this figure requires care.
On the right axis we plot the isotropic luminosity for a distance of
3.5\,kpc \citep[see][]{hcb+07}.  For comparison, the spin-down luminosity
of \axp\ in early 2007 was approximately $2\times10^{33}$\,ergs\,s$^{-1}$
\citep[see][]{ccr+07}.
}
\end{center}
\end{figure}

With the detection of \axp\ down to 2\,mm wavelength, and shortward of
$2.2\,\mu$m, there remains a factor of 1000 in wavelength where it is
undetected (Wang et al.\ 2007\nocite{wkh07} report upper limits at 24,
8, and $4.5\,\mu$m).  As discussed earlier, the radio--to--millimeter
spectrum is somewhat uncertain, and is variable, but it is flat enough
that the radio spectrum extrapolated to IR wavelengths exceeds the
IR fluxes (possibly by a very large amount; see Fig.~\ref{fig:sed}).
Therefore, it is possible that the radio spectrum steepens smoothly to
join the IR.   Contemporaneous detections in several IR bands would be
useful to delineate the shape of the spectrum in this region.  Similarly,
detection of IR pulsations from \axp\ would be of great help in pinning
down the emission mechanism, but this has not been attempted due to the
faintness of the source.

The X-ray spectra in Figure~\ref{fig:sed} represent blackbody fits to
the high and low states observed historically, discussed in more detail
by \citet{hg05}.  These are corrected for interstellar absorption,
which has been fitted as a free parameter to the X-ray spectrum.
The high-state spectrum requires two blackbodies to get a satisfactory
fit, while detailed models involving the effects of the stellar atmosphere
and resonant cyclotron scattering in the magnetosphere can reproduce the
spectrum using one surface temperature and one magnetospheric electron
temperature \citep{gol06,gogk07}.  In either case, it is evident that
the Rayleigh-Jeans tail of the thermal X-ray spectrum cannot account
for the IR flux, which must come from a separate mechanism. (Until
recently, it was customary to fit X-ray spectra of AXPs with a steep
power-law plus a blackbody, but this is unphysical because it grossly
overpredicts the optical and IR fluxes, and because Comptonization of
X-ray blackbody photons can only produce high-energy power-law tails,
not low-energy ones.)

\citet{rti+04} noted that the IR and X-ray flux from \axp\ had both
decreased by a factor of about 2 within a period of 5 months (see the
first two observations in Table~\ref{tab:ir}).  They considered that a
fossil disk reprocessing some X-rays might account for this correlation.
However, we have shown that the IR emission from \axp\ fluctuated,
rather than following a monotonic decreasing trend (\S~\ref{sec:ir}).
In 2006 September, it was a factor of about 2 fainter than when observed
by \citet{rti+04} 2.5 years before (Table~\ref{tab:ir}).  For comparison,
the X-ray flux decreased monotonically by a factor of $\sim 20$ in the
same period \citep{gh06}.  Therefore, the IR and X-ray fluxes are not
simply correlated.

Four AXPs, including \axp, have displayed IR variations, although no
general trend is evident thus far.  In 1E~2259+586, following an X-ray
outburst and accompanying IR increase, both decayed similarly and the
IR flux reached the quiescent level after $\sim 1$\,yr \citep{tkvd04}.
In 1E~1048.1--5937, large IR variations are not correlated with
the spin-down rate, but may be anticorrelated with the X-ray flux
\citep{dvk05}.  And in 4U~0142+61, although the IR, optical, and
X-ray fluxes (and spectra) all vary, including IR variations of over
a magnitude on a timescale of days, there are no clear correlations
\citep{dv06}.  In \axp, the latest IR flux is also the faintest yet
observed, and the presently observed spin-down rate is the smallest
on record \citep{ccr+07}.  The previous IR observations, brighter and
perhaps showing some variability (Table~\ref{tab:ir}), all took place
within 1.8\,yr of the original large X-ray outburst \citep{ims+04}.  It is
therefore possible to postulate that the latest IR detection corresponds
to some sort of quiescent level, much like the star has approximately
reached in X-rays \citep{gh06}, while the earlier observations reflected
at least in part transient responses from an active magnetar, related
to the outburst that led in the first place to the discovery of \axp.
However, the sampling of the IR observations is limited, and no clear
conclusion can yet be reached.


Likewise, we do not yet know whether the IR and radio fluxes are
correlated.  The original radio detection, in 2004 January, had a
flux density at 1.4\,GHz of 4.5\,mJy \citep{hgb+05}; at the time of
the latest IR observation in 2006 September, this was near 1.6\,mJy
(Table~\ref{tab:radio}); and between 2006 February and September, the flux
density ranged between 13 and 1\,mJy, with day-to-day fluctuations by
factors of up to $\sim 3$ and a general downward trend \citep{ccr+07}.
At least in 2006, therefore, the radio flux fluctuated by a greater
factor than did the IR between 2003 and 2006.  But given the limited
IR and nearly non-existent simultaneous radio--IR sampling, it remains
possible that the fluxes in these two bands may be related.  Simultaneous
IR and radio observations sampling a variety of timescales should settle
this question.

\acknowledgements

We thank Pierre Cox for approving the director's discretionary time
observations at IRAM, and the IRAM staff in Andaluc\'ia for wonderful
support during our visit.  At Gemini, we thank Jean-Ren\'e Roy for
approving a director's discretionary time proposal (GN-2006B-DD-3), and
Scott Fisher and Andrew Stephens for help with observations.  The Gemini
Observatory is operated by the Association of Universities for Research in
Astronomy, Inc., under a cooperative agreement with the National Science
Foundation (NSF) on behalf of the Gemini partnership.  The GBT and VLA
are telescopes operated by the National Radio Astronomy Observatory,
a facility of the NSF operated under cooperative agreement by Associated
Universities, Inc.  The Parkes Observatory and the ATCA are part of the
Australia Telescope, which is funded by the Commonwealth of Australia
for operation as a National Facility managed by CSIRO.  This work
was supported in part by the NSF through grant AST-05-07376 to F.C.
A.K. acknowledges financial support from the 6th European Community
Framework program through a Marie Curie, Intra-European Fellowship.

\end{document}